\documentclass[twocolumn,showpacs,preprintnumbers,amsmath,amssymb]{revtex4}

\usepackage{graphicx}
\usepackage{dcolumn}
\usepackage{bm}
\begin{document}
\title{Is turbulent mixing a self-convolution process ?}
\author{Antoine Venaille}
\email{venaille@coriolis-legi.org}
\author{Joel Sommeria}
\affiliation{Coriolis-LEGI 21 rue des martyrs 38000 Grenoble France}

\date{\today}

\begin{abstract}
Experimental results for the evolution of the probability distribution function  (PDF) of a scalar  mixed by a turbulence flow in a channel are presented. The sequence of  PDF from an initial  skewed distribution to a sharp Gaussian is found to be non universal. The route toward homogeneization depends on the ratio between the cross sections of the dye injector and the channel. In link with this observation, advantages, shortcomings and applicability of models for the PDF evolution based on a self-convolution mechanisms are discussed. 
\end{abstract}



\maketitle

Predicting the temporal evolution of the probability distribution function (PDF) of a tracer is of wide interest, in particular for the dynamics of density stratified fluids or for
reactive flows \cite{pope}. The scope of existing phenomenological  approaches is to  provide simple, efficient models to describe the evolution of the PDF, from an arbitrary initial condition to a sharp peak centred around the mean, representing the late stage of mixing.

One of the simplest and widely used models is the \emph{linear mean square estimate} (LMSE)  also called \emph{interaction by exchange with the mean} (IEM) \cite{obrien}. The main limitation of this model is that the PDF keeps its shape while it  contracts around the  mean by mixing. Many studies have been devoted to the improvement of  this approach  \cite{pope91,fox,heinz,sabel,meyer}.

There exists an alternative class of models, based on a self-convolution process. The first occurrence of such a model is the coalescence dispersion (CD) mechanism of \cite{curl}, originally intended to describe drop interactions in a two liquid system: two drops having independent scalar values $\sigma_0$ and $\sigma_0^{'}$ coalesce into one drop having the average scalar value $\sigma=(\sigma_0+\sigma_0^{'})/2$, which immediately splits into two drops with the same scalar value $\sigma$. The probability $\rho$ to measure a given scalar value after coalescence is the \emph{convolution} of the same probability function  $\rho_0$ before coalescence : $\rho\left(\sigma\right)= 2 \int \rho_0\left(2 \sigma-\sigma_0\right) \rho_0\left(\sigma_0\right) d\sigma_0$.  CD models with self-convolution have  been widely studied and extended to turbulent flows \cite{pope82} although some of the original underlying hypothesis become questionable \cite{dopazo}.
Following the ideas of the CD approach, a model of aggregation of scalar streaks has been presented in \cite{villermaux}.  A different physical mechanism has been proposed in \cite{venaille} to justify a pure self-convolution process. This model describes the temporal evolution for the \emph{coarse-grained} PDF at a given scale $l$ . The key idea is that random straining transfers scalar fluctuations toward scales smaller than $l$, so that the measured fluctuations are averaged over scale $l$. Such coarse-grained averaging is somewhat equivalent to the smoothing effect of diffusivity.

The question of the PDF evolution of a passive tracer in a turbulent flow has been addressed experimentally in different configurations (see for instance \cite{jayesh,antonia}). 
For a flow without a mean scalar gradient, convergence from a very skewed PDF toward a gaussian shape has been qualitatively observed, although some departure from gaussianity was noticed to persist even at late stages of mixing.  

 \begin{figure}
 \includegraphics[width=7cm]{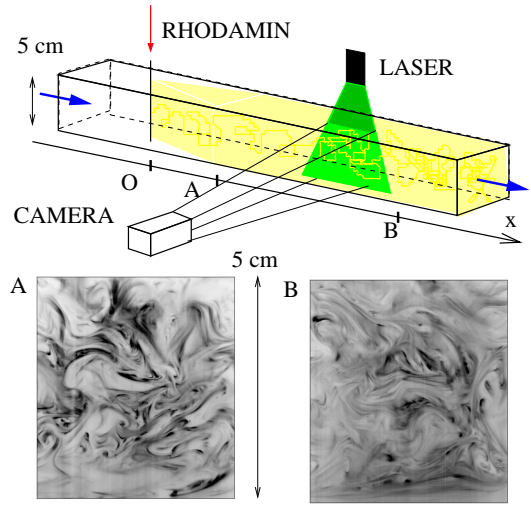}
  \caption{\footnotesize Experimental set up}
\label{Fig:set_up}
\end{figure}

Surprisingly, there seem to be only few comparisons between phenomenological models and the laboratory experiments presented above. We consider here a fluorescent dye introduced in a channel, where it is advected by a turbulent flow. PDF are measured at different locations along the channel. The PDF becomes sharper and sharper around the mean with increasing downstream distance from the injector. The sequence of PDF, from an initial skewed distribution to a sharp gaussian, is found to be non universal, depending on the injection system. Our aim is here to  discuss the capability of models based on a self-convolution process to describe those experiments.\\

\paragraph{Models}

We consider an idealised case : the scalar field and the turbulent flow are supposed to be statistically homogeneous in space.  We call $\rho_l\left(\sigma,t\right)$ the PDF of a scalar value $\sigma$ at time $t$, measured with a probe of size $l$. The scalar field measured by this probe is averaged at scale $l$ by a low pass filter. The self-convolution model relies on two hypothesis : 1) it ignores fluctuations of the straining histories of scalar sheets 2) it assumes independence of scalar values at two points separated by a distance $l$.

Hypothesis 1) implies that if the width of a scalar sheet, equal to $l$ at time $t=0$, is divided by 2 at time $t$, then the width of the adjacent sheets are also divided by a factor 2.  This time interval $t$ is related to the strain rate $s$ by the relation $\exp\left(\int_0^{t} s (t^{\prime}) dt^{\prime} \right)=2$. The probability to measure a given scalar value at scale $l$ at time $t$ is then 
  $\rho_l(\sigma,t)= 2 \int p_{l}(\sigma_1,2 \sigma-\sigma_1,0)  d\sigma_1$, where $p_l(\sigma_1,\sigma_2,0) $ is the joint probability to measure the values $\sigma_1$ and $\sigma_2$ at two points separated by the distance $l$. The hypothesis 2) of independency gives $\rho_l(\sigma,t)=2 \int \rho_{l}(\sigma_1,0) \rho_l(2\sigma-\sigma_1,0)d\sigma_1 $.

Taking the limit of continuous times for this process, a dynamical equation for the Laplace transform of the PDF $\widehat{\rho}_l(\kappa)=\int_0^{+\infty} \rho_l (\sigma) e^{-\kappa \sigma} d \sigma$  has been provided in \cite{venaille}:

\begin{equation}
\partial_t  \widehat{\rho_l} = s(t) \left[\widehat{\rho_l} \ln \widehat{\rho} -\kappa  \partial_{\kappa} \widehat{\rho_l} \right] \label{Eq:etirement_pur}
\end{equation}

Note that this equation is the continuous form of the discrete map  used in \cite{pumir}. The solution of (\ref{Eq:etirement_pur}) is

\begin{equation}
 \widehat{\rho}(\kappa,t)={\bigg [\widehat{\rho}\bigg (\frac{\kappa}{f},0\bigg )\bigg ]}^{f}  \ , \  f(t) = \exp  \left( \int_0^t s \left(t'\right)dt' \right)
\label{Eq:solution}
\end{equation}

From this expression, one can easily deduce (see \cite{venaille}) the temporal evolution of the cumulant  $c_n(t)={(-\partial_{\kappa})}^n \ln (\rho_l(\kappa,t)) \big|_{\kappa=0}$:  $c_n(t)=c_n(0)/f^{n-1}$. Cumulants are equal to the centred moments for $1 \le n \le 3$, and related to them for higher $n$. The application to the second moment yields $dc_2/dt= -s(t) c_2$. The cascade rate of variance is therefore proportional to the strain rate $s$. 

The aggregation model provided in \cite{villermaux}  leads to a different dynamical equation:

\begin{eqnarray}
\partial_t  \widehat{\rho} = s(t) \left[ f(t) \left[{\widehat{\rho}}^{1+1/f}-\widehat{\rho}\right]-\kappa  \partial_{\kappa} \widehat{\rho} \right],  \label{Eq:villermaux}
\end{eqnarray}

the solution  approaches  a sequence of gamma-PDF at large time:  $\gamma(\sigma/\langle \sigma \rangle)= \frac{f^f}{\Gamma(f)}\frac{\sigma^{f-1}}{{\langle \sigma \rangle}^{f-1}} e^{-f \sigma/\langle \sigma \rangle} $. 


\paragraph{Experimental set up}
We drive a steady flow in a water duct of square section $l_s \times l_s =5 \times 5 cm^2$, at a steady volume rate $Q$, corresponding to a bulk velocity $U=Q/(l_s^2)$. The duct is 150 cm long. Straight circular buffer sections, 50 cm long and 5 cm in diameter, are fitted at both ends to reduce perturbations from end effects. Furthermore, the upstream buffer section is filled by a honey-comb. The fluorescent tracer (Rhodamin 6G) is introduced at 50 cm downstream from the duct inlet, and observed until an abscissa $x$=80 cm downstream, in a region of well established duct flow. Most of the experiments have been performed for $U$=33 cm/s, so the corresponding Reynolds number $Re=l_s U/ \nu $ is equal to $1.6\ 10^4$ . This is about 8 times the Reynolds number of transition, so that  turbulence is well established. Test experiments performed at lower and  higher Re (still of order $10^4$) yield very similar results. 

The mean kinetic energy dissipation rate is estimated using standard measurement on turbulent channel flows \cite{schlichting} : $\epsilon= (0.316 U^3) /(2 l_s Re^{1/4}) \sim  8. 10^{-3}$ $m^2.s^{-3}$. The corresponding dissipative Kolmogorov scale is $\eta=\left(\nu^3/\epsilon \right)^{1/4} \sim 0.1$ $mm$. To estimate the Taylor microscale $\lambda=u_{rms}/<(\partial_x u)^2 >^{1/2}$, we consider that turbulent fluctuations represent $3.5\%$ of the mean velocity $U$  (following \cite{hinze}) and use the isotropic relation $\left< \left(\partial_x u\right)^2 \right>=\epsilon/(15 \nu)$. This yields $\lambda=0.5$ $mm$, and the corresponding Reynolds number is $Re_{\lambda}=u_{rms} \lambda /\nu \sim 5$. This indicates the absence of a multiscale energy cascade.
   
The Schmidt number of the tracer is $Sc=\nu/D \sim 1000$, corresponding to a very low Batchelor scale $l_b= \eta Sc^{-1/2} \sim 4 $ $\mu m$.  Two kinds of dye injectors have been used. The first one is a tapped vertical tube with eight equidistant small holes, approximating a vertical line source with uniform flux. The second is a single tube of diameter  $2 mm$, curved in the downstream direction with dye flow rate adjusted to minimise shear with the background flow. It approximates a point source at the centre of the duct. 

 Using the \emph{Laser Induced Fluorescence} technique, the scalar field is measured in a vertical plane perpendicular to the channel cross section (see figure \ref{Fig:set_up}). The PDF of concentration is measured, after proper calibration, as a histogram of images from a CCD camera (SMD), 1024x1024 pixels with 12 bit grey levels. Histograms are averaged over series of 300 images, made at a given distance $x$ downstream from the injector. We limitate the statistics to a central band, 1/4 of the duct width, where the mean concentration can be considered as uniform, and shearing by the mean flow negligible. Then the variation of the PDF with downstream distance can be assimilated as an evolution with time $t=x/U$, using the Taylor hypothesis. The spatial resolution of the images is 50 $\mu m$.  Analysis of  the scalar spectra revealed no particular power law, as well as a cut off at $k\sim l^{-1}$ which corresponds to the laser sheet thickness $l\sim 1mm$ (20 pixels). It implies that the scalar field measured is filtered at this scale $l$, which is also found to be a typical value of the scalar field correlation length.

\begin{figure}
 \includegraphics[width=6cm]{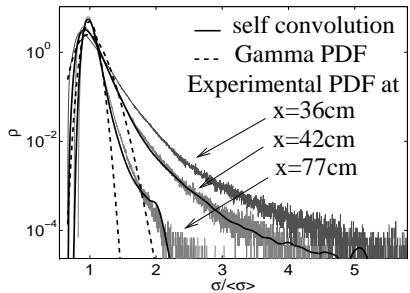}
 \caption{ \footnotesize  Line injector. The experimental PDF at $x=36cm$ and  variances at $x=42cm$ and $x=77cm$  are used for the  prediction of the self-convolution model. The fit with Gamma-distribution requires only the variance value.}
\label{Fig:comparaison}
\end{figure}

\paragraph{Mixing from a line source}
Concentration PDF measured at increasing distance from the injector are shown in figure \ref{Fig:comparaison}. It has  been compared to the prediction of the self-convolution model, shown as a solid line. For that purpose we start from the measured ``initial'' PDF at  $x=36$ cm, and apply solution (\ref{Eq:solution}), with $f(t)$ fitted to the measured variance $c_2(t)$. Strikingly, there is good agreement with the model based on a pure convolution mechanism. We see no tendency for the PDF to approach  \emph{gamma-PDF} (dashed line).


\begin{figure}
\includegraphics[width=6cm]{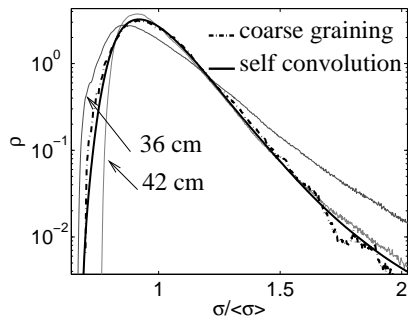}
 \caption{ \footnotesize  Line injector. Experimental PDF at $x=36cm$ and $x=42cm$. The coarse graining is applied to the scalar field at $x=36cm$.}
\label{Fig:comparaison_cg}
\end{figure}

In figure \ref{Fig:comparaison}, the initial PDF, $\rho_{l}(x_0)$ has been obtained by analysing scalar fields at the resolution of the laser sheet. What would happen if the scalar field were coarse-grained with a filtering function of length $l'>l$  ? The PDF $\rho_{l'}(x_0)$ is expected to become sharper and sharper around its mean value, as if the scalar field were ``mixed''. This filtering process therefore has the qualitative effect of a diffusivity. 
The initial PDF $\rho_{l}(x_0)$ is reproduced in figure \ref{Fig:comparaison_cg}. Let $c_2(0)$ be its variance. We then  compare the coarse-grained PDF $\rho_{l'}(x_0)$ with  the PDF $\rho_{l}(x)$ measured at a distance $x$ such that their variances have the common value $c_2=c_2(0)/2$. We also plot the self-convolution of the initial PDF $\rho_{l}(x_0)$. 
The PDF shown are zoomed close to the mean value because the filtering process reduces the number of independent events for the construction of the PDF $\rho_{l}(x_0)$.  As expected, the coarse-grained evolution is qualitatively close to the temporal evolution, with discrepancy for low scalar value. In addition, the self-convolution of the initial PDF is in very good agreement with the coarse-grained PDF.

\begin{figure}
\includegraphics[width=6cm]{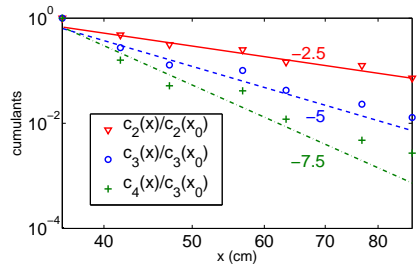}
  \caption{\footnotesize Line injector. Evolution of the first cumulants. The dashed and dot-dashed line are prediction of the self-convolution model, by supposing $f \sim x^{\alpha}$, where $\alpha=2.5$ is obtained from the  experimental variance decay} \label{Fig:moments}
\end{figure}

The comparison between the decay of different cumulants of the experimental PDF $\rho_l$  with the prediction of the model reveals a limit of the self-convolution approach.  On figure \ref{Fig:moments}), we fit the variance decay with a power law $c_2(x)\sim x^{-2.5}$, which is consistent with experimental results of \cite{villermaux}. The self-convolution model predicts then a decay $c_3 \sim x^{-5}$ and $c_4 \sim x^{-7.5}$. We notice fluctuations of experimental data around this prediction, and a slight overestimation of the actual decay of the cumulants, growing at increasing distance from the injector.

\begin{figure}
\includegraphics[width=6cm]{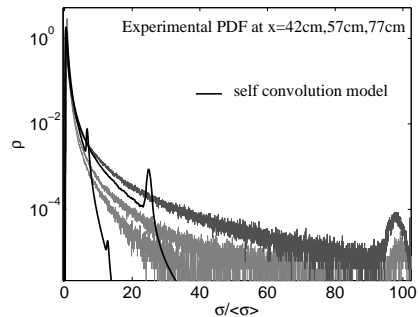} 
\caption{ \footnotesize Point injector. The fit with the model of pure self-convolution is done by taking the first measured PDF (dark grey) as the initial condition.}
\label{Fig:inj_ponct}
\end{figure}

\paragraph{Mixing from a point source}
The sequence of experimental  PDF presented in figure \ref{Fig:inj_ponct} has a characteristic shape that indicates that almost no mixing has occurred.  There are two sharp peaks :  one is the backround flow, with a small mixed scalar concentration, and the other one corresponds to the existence of unmixed blobs of scalar. This second peak persits in the early evolution of the distribution. Then the peak disappears, but the PDF remains very skewed, showing no convergence toward a \emph{gamma-PDF}. It is also clear that the model (\ref{Eq:etirement_pur}) does not provide good prediction in this case: there is a spurious peak. 

\paragraph{Discussion}
We observe that the self convolution model provides good predictions in the case of the line injector, but discrepancy clearly appears with the point injector. We here discuss the difference between these two cases in relation with our initial hypothesis. Since the capability of the model to describe the time evolution of the PDF depends on the scalar injection  in the same turbulent flow, the observed descrepancy should not be sought in the fluctuations of $f(t)$ (first model hypothesis).

 To discuss the validity of the second hypothesis, i.e. the independence of scalar probability functions, let us consider a one dimensional idealization of the injection system: the scalar  field is supposed to be a succession of voids and colored  segments, uncorrelated to each other. While the corresponding PDF is  a double Dirac function in both cases, the PDF of the coarse grained fields may be very different. Two coarse grained scalar concentrations, separated by a distance $l$, can be considered independent only if $l \gg \max\{\overline{l}_0(t), \overline{l}_{\sigma}(t) \}$, where .$\overline{l}_0(t)$ and $\overline{l}_{\sigma}(t)$ are typical lengths of voids and colored segments. For both the line and the point case, we are in a dilute limit: $\overline{l}_0(t) \gg \overline{l}_{\sigma}(t)$. As seen in figure 6-b, in the case of a point injector, the voids length scale $\overline{l}_0 $ is larger than the probe length scale $l$ (20 pixels). It has been estimated more quantitatively  to be around 80 pixels by computing the  statistics of voids. It is thus not surprising that the self-convolution model does not work in that case. By contrast, as seen in figure 6-a, it is not possible to find void events in the coarse grained scalar field. It indicates that the typical void length scale is smaller than the probe length scale $l$.

\begin{figure}
\includegraphics[width=7.5cm]{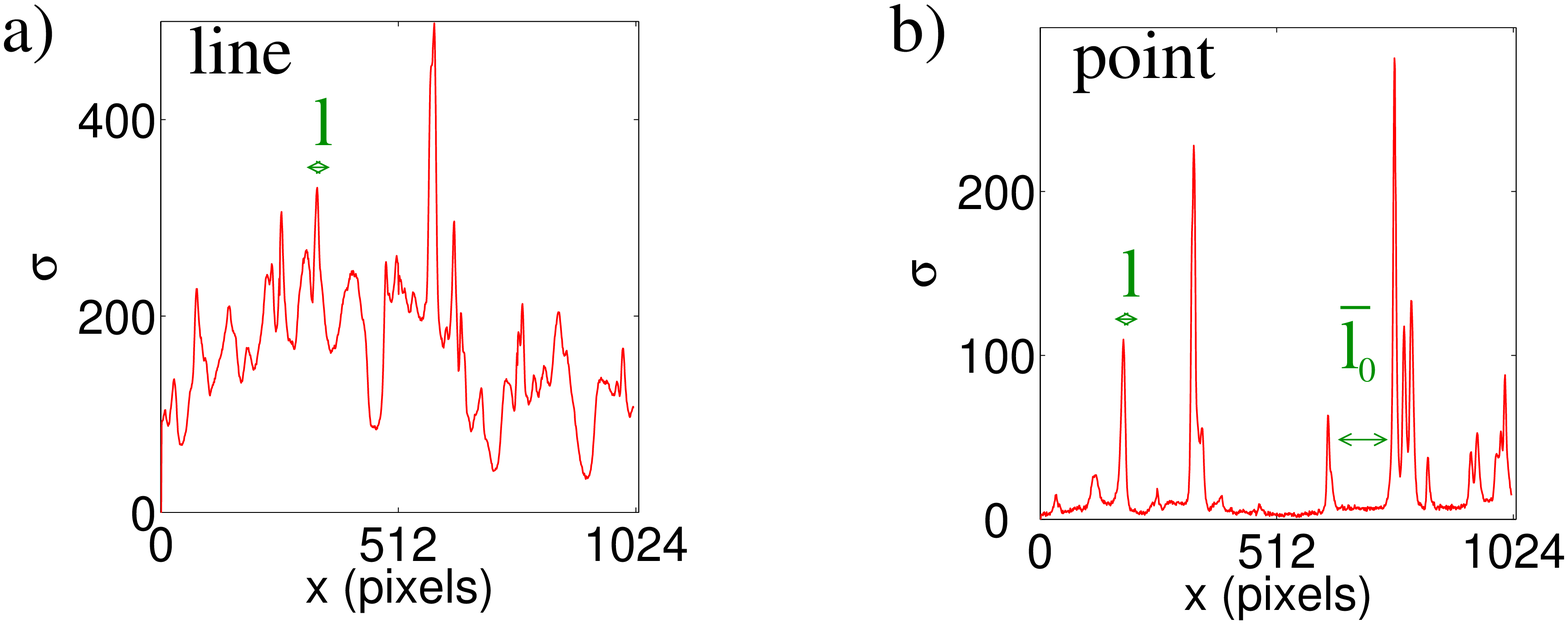}
\caption{ \footnotesize Typical scalar profile on an image taken at 40 cm from the injector. The total length is 5 cm (1024 pixels).  The filtering length scale l (laser sheet thikness) is around 20 pixels.}
\label{Fig:line_field}
\end{figure}

More generally, those length scales are related to the channel and the injector properties. The ratio of their cross section surface, respectively  $\sim r l_s/ l_s^2$ and $\sim r^2/l_s^2$ for the line and the point case, is equal to the mean scalar value  $\left<\sigma\right>=\overline{l}_{\sigma}(t) / (\overline{l}_0(t)+\overline{l}_{\sigma}(t) )$. The distance $r \sim 2$  mm is the initial scalar sheet thickness (either the width of the vertical tube or the diameter of the small cylinder). This length is then reduced by straining effects $\overline{l}_{\sigma}(t)=r/f(t)$, and  the random succession of scalar sheets create void segments, while conserving $\left<\sigma\right>$. This gives an estimation of the void length scale:   $\overline{l}_0(t)=l_s/f(t) $ for the line injector, and  $\overline{l}_0(t)=l_s^2 / (r f(t)) $ for the point injector. As far as $r\ll l_s$, we see that void length scales differ by an order of magnitude. This is consistent with experimental observation: voids events are of the order of a few filtering scales $l$ for the point injector, and smaller than $l$ for the vertical injector. Note that at sufficiently large time, the condition for independence of scalar sheets will always be realized.

\paragraph{Conclusion}
 If a gamma PDF is initially created, our self-convolution model predicts that it will evolve toward a Gaussian through a succession of gamma-PDF, in exact agreement with \cite{villermaux}. Our experiments however provide no evidence for a general convergence of scalar PDF toward gamma pdf, in contrast with the model prediction of \cite{villermaux}. Further comparisons of different injection systems would be needed to confirm this point.

In the case of a line injector, our experiments provide good support of the self-convolution model. In this model, the evolution of the scalar PDF by the straining effect is equivalent to coarse-graining. We have directly checked this equivalence in the experiments.For practical purposes, this self-convolution model has two interesting  properties: i) it conserves scalar bounds and mean scalar value ii) it evolves toward a sharp gaussian at large time, by contrast with the LMSE (or IEM) model.  

The validity of the self-convolution model depends on the ratio between  the cross sections of the  injector and the channel. This ratio prescribes length scales for the fine grained scalar structures (length of colored and void segments ), which decrease with time by straining effects. The model is correct when the probe length scale $l$ is larger than the typical scalar field length scales, to assure independence between scalar sheets at distance $l$. 

The model also relies on the hypothesis of weak fluctuations for the straining rate. This is typically expected for the Batchelor regime of scalar cascade or for moderate Reynolds numbers, as in our experiments. Discrepancies are expected to arise from intermittency in a well developped Kolmogoroff cascade.

\paragraph*{Aknowledgment}
\small H. Didelle and S. Viboud have been of great help for the set up of the experiment.We thank A. Gagnaire and P. Kramer for their participation in the experiments and data processing.

\bibliography{mixing}

\end{document}